\documentstyle[manuscript,osa]{revtex}
\begin{document}
\begin{center}
{\large \bf Angular momentum transfer to a star by gravitational waves}\\
{H. G. Khosroshahi$^1$ and Y. Sobouti$^{1,2}$}\\
{\small $^1$Institute for Advanced Studies in Basic Sciences, P. O. Box 45195-159,
Zanjan, Iran}\\
{\small $^2$Physics Department, Shiraz University, Shiraz, Iran}\\
{\small email: khosro@sultan.iasbs.ac.ir,  sobouti@sultan.iasbs.ac.ir}
\end{center}
\noindent
{ \bf Summary and Introduction}\\
Interaction of a stochastic background of gravitational radiation with celestial systems
changes their dynamical elements in a random manner and give rise to secular changes
in time(Berttoti 1973, Mashhoon $etal$ 1981, Khosroshahi and Sobouti 1997).
It has been speculated that a study of such secular changes might serve as a possible
mean of detecting gravitational radiation. In this spirit we study the angular momentum
transfer from a random background  of radiation either to a rotating star or to an oscillating
one. The angular momentum transferred to such objects by a continuous plane wave
is proportional to time, $t$, and by an stochastic background is proportional to $t^{1/2}$.\\
\newpage
\noindent
{ \bf 1. Interaction of a rotating star with gravitational waves}\\
Consider a polytrope rotating with an angular frequency $\Omega$, about
the z-axis. The density $\rho$ at  position ${\bf r}$, may be written as
\begin{equation}
\rho({\bf r},b)=\rho_0(r)+b[\rho_1(r)+\rho_2(r) P_2(\cos \theta)]
\end{equation}   
where $b=\Omega^2/\omega_{osc}^2$ is an expansion parameter and 
$ \omega_{osc}=\sqrt{4\pi G\rho(0)} $ is of the order of the angular frequency of natural oscillations
of the star.
Suppose that a monochromatic plane gravitational wave with frequency $\omega$ falls
on the star in a direction characterized by azimutal and  polar angles $\alpha$ and $\beta$,
respectively. Assume the radius of the star is much smaller than the wavelength of the incident 
wave. In a Fermi coordinate system, the tidal force on unit mass exerted by the  wave  is the
gradient of a scalar,
\begin{equation}
{\bf f}^{gw}=- \nabla\Phi=\nabla[\frac{1}{4}\omega^2\epsilon\sum_{i,j=1}^3x_ih_{ij}(t)x_j],
\end{equation}
where $\epsilon\ll 1$ is a certain characteristic amplitude of the wave and is independent of the 
frequency $\omega$ and  $h_{ij}(t)=Re \{A_{ij}(\omega)exp(-i\omega t+i{\bf k}.{\bf x})\}$
are the spatial components of fluctuations in the space-time metric by the incident 
wave. $h_{ij}$  can be taken to be transverse and traceless. 
The torque exerted on the star, measured by a corotating observer, is
\begin{mathletters}
\begin{equation}
\vec\tau=\frac{d {\bf L}}{dt}=\int_v {\bf r}\times\rho{\bf f} ^{gw} d^3{\bf r}
=\frac{\pi}{30}
\frac{MR^2}{\bar\rho} b\epsilon A(\omega)\omega^2 \int_0^1 \rho_2(x)x^4 dx
 \{h_1(t) {\hat i}+h_2(t){\hat j} \}
\end{equation}
where ${\bar \rho}, M, R $ are the mean density, mass and radius of the star, respectively,
\begin{equation}
h_1(t)=\frac{1}{2}\sin 2\beta \sin (\Omega t + \alpha)\cos \omega t -\sigma \sin \beta 
\cos (\Omega t +\alpha) \sin \omega t, 
\end{equation}
\begin{equation}
h_2(t)=\frac{1}{2}\sin 2\beta \cos (\Omega t + \alpha)\cos \omega t +\sigma \sin \beta 
\sin (\Omega t +\alpha) \sin \omega t,
\end{equation}
\end{mathletters}
and $\sigma=1$, and $-1$ for right and left circular polarization, respectively.
We substitute eqs.(3b,c) in (3a) and integrate with respect to time. In resonance the
secular change in the angular momentum is proportional to $t$. Furthermore, this change 
is in the xy-plane and manifests itself as a precession of the rotation axis.
In an isotropic background of radiation, the mean precession
amplitude is zero. Its root mean square value, however, changes as $t^{1/2}$. Thus,   
\begin{mathletters}
\begin{equation}
\theta_{rms}^2\simeq \frac{L_{xy}^2}{L_z^2}\simeq\frac{1}{L_z^2}\int d(\cos\beta) d\alpha
[\int_0^t {\vec \tau} dt]^2 d\omega,
\end{equation}
and finally,
\begin{equation}
\theta_{rms}^2 \sim \frac{\pi G b^2}{\bar \rho}[\int_0^1 \rho_{b2}(x)x^4 dx]^2
{\cal S}_g(\Omega)t~~~~~~~~~~for~~t\gg\Omega^{-1},
\end{equation}
\end{mathletters}
\noindent
where ${\cal S}_g(\omega)=\epsilon^2A(\omega)^2\omega^2/8\pi G$ is the energy spectral density
of the background  radiation.
Equation(4) indicates a random walk for the rotation axis of the star.\\\\
\noindent
{\bf 2. Interaction with normal modes of a star}\\
The torque of a gravitational wave on a rotating star, eqs.(3), is due to deviation from the
spherical symmetry of the star. This asymmetry may come about because of the natural oscillation of the star. Helioseismic waves in different spherical harmonic numbers are
prominent examples of such occurrence. This is the justification for the following
analysis .\\  
Let $\rho(r)$, $p(r)$ and $U(r)$ denote the density, the pressure and 
the gravitational potential of a star in hydrostatic equilibrium. Let a mass element
at {\bf r} undergo an infinitesimal displacement $\xi({\bf r},t)$
from its equilibrium position. It causes small changes  $\delta\rho({\bf r},t),\delta  p({\bf r},t)$
and $\delta U({\bf r},t)$. The linearized Euler's equation of motion is  
\begin{equation}
-\rho\ddot\xi=\nabla(\delta p)+\delta\rho\nabla\Omega+
\rho\nabla(\delta U)={\cal W} \xi~ , 
\end{equation}
where,
\begin{mathletters}
\begin{equation}
\delta\rho=-\nabla.(\rho\xi),
\end{equation}
\begin{equation}
\delta p=\frac{dp}{d\rho}\delta\rho-[(\frac{\partial p}{\partial \rho})_{ad}
-\frac{dp}{d\rho}]\rho\nabla.\xi, 
\end{equation}
\begin{equation}
\nabla^2(\delta U)=-4\pi G \delta\rho.
\end{equation}
\end{mathletters}
The displacement $\xi$ belongs to a function space ${\cal H}$ in which the inner product is defined as
$(\eta ,\rho\xi)=\int\rho\eta^*.\xi~d^3x=finite, ~~~\xi,\eta \in {\cal H}.$
The operator ${\cal W}$ is self-adjoint on ${\cal H}$ and gives rise to the eigenvalue problem
${\cal W}\xi_{n}=\omega_{n}^{2}\rho\xi_{n},$ where $\omega_{n}^{2}$ are real.
Using a gauged version of Helmholtz's theorem, one may decompose a general vector into 
an irrotational and a ``weighted'' solenoidal component. Thus
\begin{mathletters}
\begin{equation}
\xi=\xi_p+\xi_g,
\end{equation}
where
\begin{equation}
\xi_p=-\nabla\chi_p;~~~~~~~~~~~~~~~~~~~~with~~\nabla\times\xi_p=0,
\end{equation}
\begin{equation}
\xi_g=\rho^{-1}\nabla\times\nabla\times ({\hat{\bf r}}\chi_g);~~~~~with~~\nabla . (\rho\xi_g)=0.
\end{equation}
\end{mathletters}
Here ${\hat{\bf r}}$ is the unit vector in ${\bf r}$ direction, and $\chi_p$ and $\chi_g$ are two
scalars. See Sobouti(1980) for details of eqs.(5-7). As in eqs.(3) the torque exerted by a
gravitational wave incident on a star is,
\begin{equation}
\frac{d{\bf L}}{dt}=\int_v \delta\rho{\bf r}\times {\bf f}^{gw} d^3 x.
\end{equation}
From eqs.(6a) and (7), $\delta \rho$ for $\xi_g$ is zero and for $\xi_p$ is
\begin{equation}
\delta\rho=\rho(r)\{\chi''_p+(\frac{2}{r}+\frac{\rho'}{\rho})\chi'_p-\frac{l(l+1)}{r^2}\chi_p\}
Y_{lm}(\theta,\phi)e^{-i\omega_n t}.
\end{equation}
Only $l=2$ modes will contribute to the torque and for simplicity we will consider
the $m=0$ case.
Equation(8) reduces to an expression similar to that of eqs.(3),  
\begin{mathletters}
\begin{equation}
\frac{d{\bf L}}{dt}=-\frac{1}{9}\frac{MR^2}{\bar\rho}A(\omega) \omega^2 \int_0^1 \rho(x)
\{\chi_p''+(\frac{2}{x}+\frac{\rho'}{\rho})\chi_p'-\frac{6}{x^2}\chi_p\}x^4 dx\nonumber
\times\{h'_1(t) {\hat i}+h'_2(t){\hat j} \}\cos\omega_n t,
\end{equation}
\noindent
where
\begin{equation}
h'_1(t)=-\frac{1}{2}\sin 2\beta \sin\alpha \cos\omega t +
\sigma\sin\beta\cos\alpha\sin\omega t 
\end{equation}
\begin{equation}
h'_2(t)=-\frac{1}{2}\sin 2\beta \sin\alpha \cos\omega t -
\sigma\sin\beta\cos\alpha\sin\omega t 
\end{equation}
\end{mathletters}
The root mean square of the amplitude of precession induced by an isotropic background of
radiation becomes
\begin{equation}
\theta_{rms}^2\sim \frac{\pi}{\bar \rho}[  \int_0^1 \rho(x) \{\chi''+(\frac{2}{x}+\frac{\rho'}{\rho})\chi'-
\frac{6}{x^2}\chi\}x^4 dx]^2{\cal S}_g(\omega_n) t 
\end{equation}
$$for ~~t\gg\omega_n^{-1}$$
A numerical evaluation of eq.(11) involves the following steps.\\
a) One calculates the $p$ and $g$
modes of the model star belonging to $l=2$. A $p-$mode will have a large
irrotational component as indicated in eqs(1). For a $g-$mode the situation will be the
opposite.\\
b) One extracts the irrotational component, $\chi_p$ of each mode, substitutes in eq.(11)
and carries out the integration numerically.
Numerical evaluations of the rms precession amplitudes are in progress.\\\\
\\\noindent
{\bf References}\\
Berttoti B., 1973, $ Ap.J.$ Lett. {\bf 14}, 51.\\
Khosroshahi H. G. , Sobouti Y., 1997, $A \& A$ {\bf 321},1024.\\
Mashhoon B., Carr B. J., Hu B. L., 1981,  $Ap.J.$ {\bf 246}, 569.\\ 
Sobouti Y., 1981, $A\& A$ {\bf 100}, 319.\\
\end{document}